\documentclass[pra,aps,twocolumn,longbibliography, superscriptaddress,groupedaddress]{revtex4-2}

\usepackage[paperwidth=210mm,paperheight=297mm,centering,hmargin=1.65cm,vmargin=2.1cm]{geometry}

\pdfoutput=1

\date{\today}

\usepackage{fourier}
\usepackage{amsmath,amssymb}
\usepackage{mathtools}
\usepackage{amsthm}
\usepackage{empheq}
\usepackage{cases}

\usepackage{times}
\usepackage{upgreek}
\usepackage{psfrag} 
\usepackage{latexsym} 
\usepackage{amstext}
\usepackage{amsfonts}
\usepackage{amsxtra} 
\usepackage{dcolumn}
\usepackage{dcolumn}
\usepackage{textcomp}
\usepackage{graphicx}
\usepackage{bm}
\usepackage{color}
\usepackage{braket}
\usepackage{units}
\usepackage[svgnames]{xcolor}

\definecolor{myColor}{rgb}{0.02,0.12,0.3}
\definecolor{myciteColor}{rgb}{0.39,0.7,0.89}
\usepackage[colorlinks=true,citecolor=myColor,linkcolor=myColor,urlcolor=myColor]{hyperref}

\def\be{\begin{equation}}
\def\ee{\end{equation}}

\def\nobreakbefore{%
  \relax\ifvmode\else
    \ifhmode
      \ifdim\lastskip > 0pt\relax
        \unskip\nobreakspace
      \else 
        \nobreakspace
      \fi
    \fi
  \fi
}
\let\oldcite\cite
\renewcommand\cite{\nobreakbefore\oldcite}

\newcommand{\kc}{k_{\textrm c}}
\newcommand{\Ec}{E_{\textrm c}}
\newcommand{\Ds}{D_{\textrm s}}
\newcommand{\Ks}{K_{\textrm s}}

\newcommand{\Dd}{D_{\textrm d}}
\newcommand{\VD}{V_{\textrm D}}

\newcommand{\kB}{k_{\textrm{B}}}

\usepackage{graphicx}
\usepackage{dcolumn}
\usepackage{bm}
\usepackage{xcolor}
\usepackage{svg}
\usepackage[titletoc]{appendix}
\usepackage{pdfpages}

\makeatletter
\let\frontmatter@footnote@produce\frontmatter@footnote@produce@endnote
\makeatother

\makeatletter
\AtBeginDocument{\let\LS@rot\@undefined}
\makeatother

\makeatletter
\newsavebox{\@brx}
\newcommand{\llangle}[1][]{\savebox{\@brx}{\(\m@th{#1\langle}\)}%
  \mathopen{\copy\@brx\kern-0.5\wd\@brx\usebox{\@brx}}}
\newcommand{\rrangle}[1][]{\savebox{\@brx}{\(\m@th{#1\rangle}\)}%
  \mathclose{\copy\@brx\kern-0.5\wd\@brx\usebox{\@brx}}}
\makeatother

\begin{document}

\title{
Energy-space random walk in a driven disordered Bose gas
}

\author{Yansheng Zhang}
\email{yz661@cam.ac.uk}
\author{Gevorg Martirosyan}
\author{Christopher J. Ho}
\author{Ji\v{r}\'{i}~Etrych}
\author{Christoph~Eigen}
\email{ce330@cam.ac.uk}
\author{Zoran Hadzibabic}
\affiliation{
Cavendish Laboratory, University of Cambridge, J. J. Thomson Avenue, Cambridge CB3 0HE, United Kingdom}
\date{\today}

\begin{abstract}
Motivated by the experimental observation \cite{Martirosyan:2023} that driving a non-interacting Bose gas in a 3D box with weak disorder leads to power-law energy growth, $E \propto t^{\eta}$ with $\eta=0.46(2)$, and compressed-exponential momentum distributions that show dynamic scaling, we perform systematic numerical and analytical studies of this system. Schr{\"o}dinger-equation simulations reveal a crossover from $\eta \approx 0.5$ to $\eta \approx 0.4$ with increasing disorder strength, hinting at the existence of two different dynamical regimes. We present a semi-classical model that captures the simulation results and allows an understanding of the dynamics in terms of an energy-space random walk, from which a crossover from $E \propto t^{1/2}$ to $E \propto t^{2/5}$ scaling is analytically obtained. The two limits correspond to the random walk being limited by the rate of the elastic disorder-induced scattering or the rate at which the drive can change the system's energy. Our results provide the theoretical foundation for further experiments.
\end{abstract}

\maketitle

\section{Introduction}

The emergence of simple and universal behaviors insensitive to system parameters and past trajectories is one of the most fascinating aspects of the physics of complex systems. Although the theory of universal behaviors was traditionally developed for equilibrium critical phenomena~\cite{Kardar:2007}, recent experimental and theoretical studies have extended these ideas to a wide range of far-from-equilibrium systems~\cite{Nakayama:1994, HalpinHealy:1995, Odor:2004, Polkovnikov:2011, Tauber:2014, Altman:2015, Langen:2015b, Munoz:2018, Mikheev:2023}. 

In particular, a broad range of universal dynamics has been observed in quenched or driven ultracold atomic gases (see, 
\emph{e.g.},~\cite{Sagi:2012,Hung:2013,Makotyn:2014,Navon:2016,Pruefer:2018, Eigen:2018, Erne:2018,Johnstone:2019,SaintJalm:2019,Glidden:2021,Galka:2022,Wei:2022,Le:2023,Huh:2023}).
One fruitful avenue for such studies involves driven box-trapped Bose gases~\cite{Navon:2016}, where the interplay of the drive and the inter-particle interactions leads to turbulent cascades with power-law momentum distributions~\cite{Navon:2016} sustained by a constant momentum-space energy flux~\cite{Navon:2019}.
While interactions are usually central to the universal dynamics, in a recent experiment \cite{Martirosyan:2023}, we demonstrate that in absence of interactions,
an interplay between drive and disorder can also lead to universal behavior. This system, with a power-law energy growth [$E \propto t^{\eta}$ with $\eta=0.46(2)$] and self-similar momentum distributions well characterised by a compressed exponential, shows qualitatively different behavior from its interacting counterpart.

In Ref.~\cite{Martirosyan:2023}, these observations are reproduced with Schr\"{o}dinger-equation simulations and qualitatively explained by a semi-classical model. In this paper, we formalize our theoretical results. First, we extend the Schr\"{o}dinger-equation simulations to a wider parameter range and observe a crossover from $\eta \approx 0.5$ to $\eta \approx 0.4$ with increasing disorder strength (Section~\ref{sec:Schrodinger_simulations}), which hints at the existence of two distinct dynamical regimes.
We then present the semi-classical model (Section~\ref{sec:Semiclassical_model}) that captures the simulation results and allows an understanding of the dynamics in terms of an energy-space random walk. This in turn leads to a simple energy-space drift-diffusion equation 
(Section~\ref{sec:Analytic_analysis}) that reproduces the crossover between the two regimes, and analytic predictions of $E \propto t^{1/2}$ and $E \propto t^{2/5}$ that emerge in the limits where the random walk is limited by the rate of disorder-induced scattering or the rate at which the drive can change the system's energy. Our results offer a new example of a dynamical system undergoing energy-space drift-diffusion~\cite{Jarzynski:1993a, Jarzynski:1993b, Bunin:2011, Hodson:2021, Hodson:2021b} and provide the theoretical foundation for further experimental studies.


\section{Schr{\"o}dinger-equation simulations}
\label{sec:Schrodinger_simulations}

\begin{figure*}[t]
    \centering
    \includegraphics[width=\textwidth]{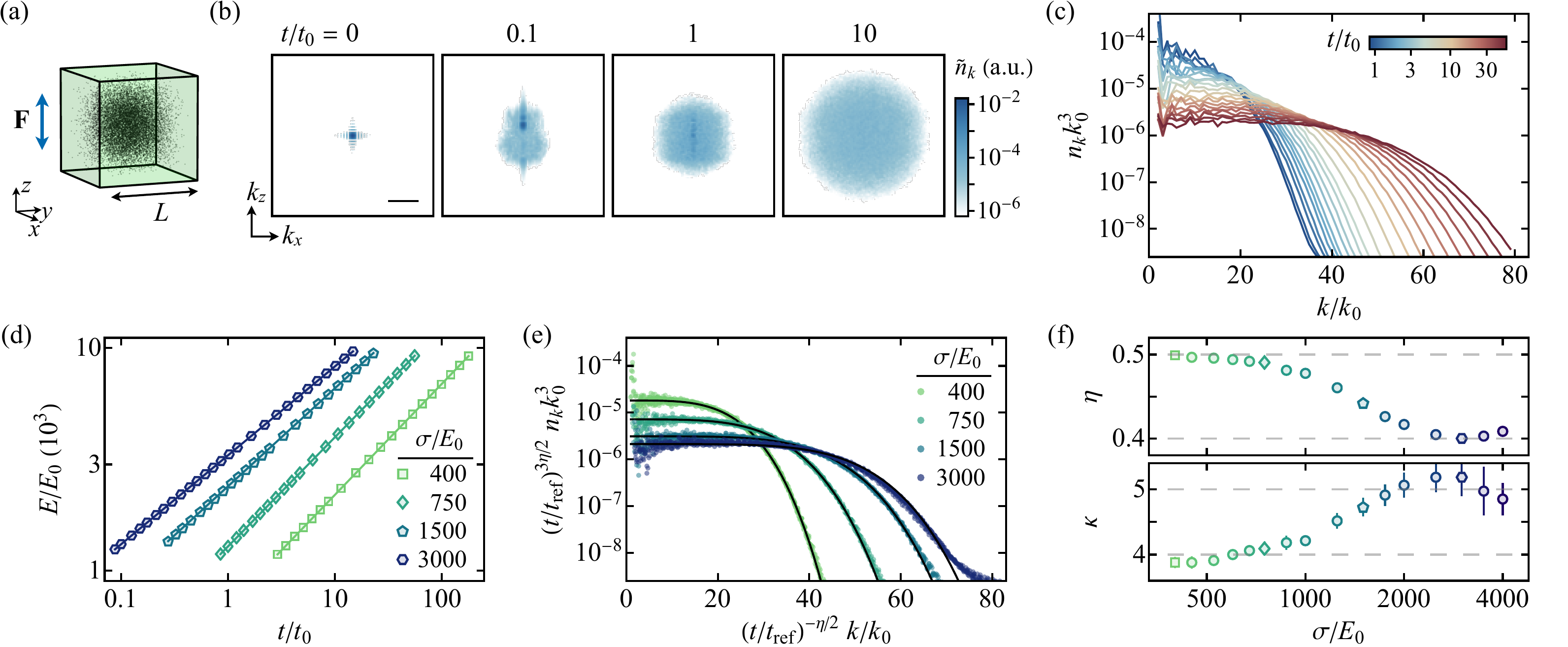}
    \caption{Schr{\"o}dinger-equation simulations of a driven non-interacting Bose gas in a box with disorder. 
    (a) Illustration of the simulation geometry. For the box of size $L$, the natural units of momentum, energy, and time are, respectively, $\hbar k_0=\hbar \pi/L$, $E_0 = \hbar^2/\left(mL^2\right)$, and $t_0 = \hbar/E_0$. (b) Snapshots of the projected momentum distribution $\tilde{n}_k(k_x,k_z)$ for drive parameters $U = 1500 \, E_0$ and $\omega = 75 \, E_0/\hbar$, and disorder strength $\sigma = \langle \VD^2 \rangle^{1/2}= 750 \, E_0$ (the simulation grid is of size $127\times127\times127$, which leads to a UV cutoff of $127 k_0$). For comparison to Ref.~\cite{Martirosyan:2023}, using the $^{39}$K atom mass $m = 6.5\times 10^{-26}\,$kg and $L = 50\,\upmu$m, the simulation parameters here correspond to $U/\kB = 7.4$\,nK, $\omega/(2\pi) = 9.5$\,Hz, and $\sigma/\kB = 3.7$\,nK. The scale bar corresponds to $20\,k_0$.
    (c) Evolution of the (spherically averaged) momentum distribution $n_k$ for parameters as in (b) and $t/t_0\in[0.85, 55.8]$.
    (d) Energy-growth dynamics for $U = 1500 \, E_0$, $\omega = 75 \, E_0/\hbar$, and various $\sigma$.
    The solid lines show power-law fits used to extract the energy-time scaling exponent $\eta$. 
    (e)  For each $\sigma$ value in (d), momentum distributions at different $t$ (such as shown in (c) for $\sigma = 750\,E_0$) collapse onto a single curve when dynamically scaled according to Eq.~(\ref{eq:dynamicscaling}), with arbitrarily chosen $t_\text{ref} = 10\,t_0$.  
    The solid lines show fits according to Eq.~(\ref{eq:compressed-exponential-def}), used to extract the compressed-exponential exponents $\kappa$. 
    (f) Extracted $\eta$ and $\kappa$ as a function of  $\sigma$ for fixed $U=1500\,E_0$ and $\omega = 75\,E_0/\hbar$.}
    \label{fig:1}
\end{figure*}

The non-interacting dynamics in Ref.~\cite{Martirosyan:2023} can be described by the Schr{\"o}dinger equation
\begin{equation}
    i\hbar \frac{\partial}{\partial t} \psi = \left[-\frac{\hbar^2}{2m} \nabla^2 + V_{\rm box} + \VD - \frac{U z}{L} \sin(\omega t)\right]\psi
    \label{eq:Schrodinger},
\end{equation}
where $m$ is the particle mass, $V_{\rm box}$ is the clean trapping potential, $\VD$ is the disorder, $L$ is the box length along the driving-force direction $\bf{z}$, and $U/L$ is the amplitude of the driving force.
Here, we model the trap as a cubic box [Fig.~\ref{fig:1}(a)] of infinite depth\footnote{In Ref.~\cite{Martirosyan:2023}, a cylindrical box trap was used, but the essential physics should be the same as long as the dynamics of the driven direction remain separable for $\VD = 0$.}, and the disorder $\VD$ is chosen to be an uncorrelated (zero-mean) Gaussian random potential. The choice of an uncorrelated potential is sensible because the correlation length of $\VD$ in an optical trap, which is on the order of the laser wavelength $\lambda$, is small compared to the atomic de-Broglie wavelength in the experiment. The strength of the  random potential $\VD$ is characterized by its r.m.s.~value~$\sigma$.

In Fig.~\ref{fig:1}(b), we illustrate the evolution of the momentum distribution, $n_k(\mathbf{k}) = |{\psi}(\mathbf{k})|^2$, for one choice of parameters \mbox{$\{U$, $\omega$, $\sigma\}$}. As in the experiment~\cite{Martirosyan:2023}, the drive rapidly increases the momentum spread along $\bf{z}$, and cross-dimensional coupling due to $\VD$ causes energy to leak into the transverse directions. At long times, $n_k(\mathbf{k})$ is nearly isotropic and gradually broadens [Fig.~\ref{fig:1}(c)]. 

In agreement with the observations in Ref.~\cite{Martirosyan:2023}, the energy growth is well described by a power-law, $E(t) \propto t^\eta$, as shown in Fig.~\ref{fig:1}(d).
The (nearly-)isotropic momentum distributions $n_k(k, t)$ at different $t$ are self-similar, with
\begin{equation}
    n_k(k,t) = \left(\frac{t}{t_\text{ref}}\right)^{\alpha} n_k\left(\left(\frac{t}{t_\text{ref}}\right)^{\beta}k , t_\text{ref}\right)\,,
    \label{eq:dynamicscaling}
\end{equation}
where $t_\text{ref}$ is an arbitrary reference time, $\beta=-\eta/2$, and $\alpha=3\beta$ corresponding to particle-conserving transport\footnote{In Ref.~\cite{Martirosyan:2023}, line-of-sight integrated distributions were analysed experimentally, in which case $\alpha = 2\beta$, with $\beta$ still equal to $-\eta/2$.}.
This self-similarity is illustrated in Fig.~\ref{fig:1}(e) for different parameters $\{U, \omega, \sigma\}$; for each simulation, the distributions at different $t$ collapse onto a single curve when rescaled according to Eq.~(\ref{eq:dynamicscaling}).
The collapsed curves are well described by compressed exponentials [black lines in Fig.~\ref{fig:1}(e)] of the form
\begin{equation}
    n_k(k) \propto \exp\left[-(k/k_{\rm s})^\kappa\right],
    \label{eq:compressed-exponential-def}
\end{equation}
with exponent $\kappa$ and momentum scale $k_{\rm s} \propto \sqrt{E}$.

In Ref.~\cite{Martirosyan:2023}, only a relatively narrow range of $\eta$ and $\kappa$ was observed. Here, by extending the range of disorder strengths~$\sigma$, we observe a crossover from $\eta\approx 0.5$ and $\kappa\approx 4$ to $\eta\approx 0.4$ and $\kappa\approx 5$ [Fig.~\ref{fig:1}(f)].
This hints at the existence of two distinct dynamical regimes, corresponding to weak and strong disorder. Analytically understanding the emergence of these two regimes is the goal of the subsequent sections.

\begin{figure*}[t]
    \centering
    \includegraphics[width=\textwidth]{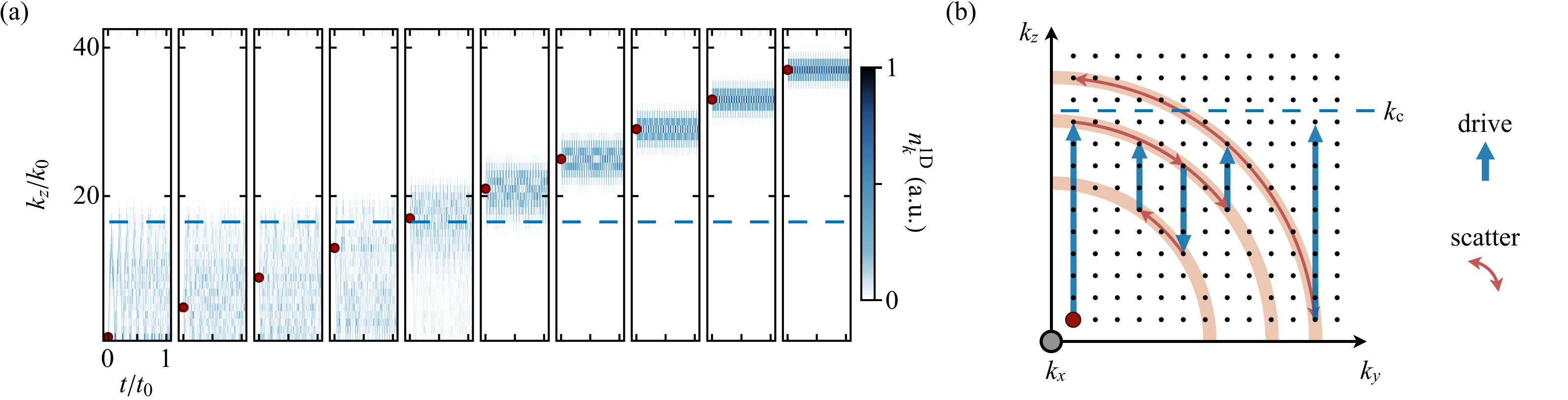}
    \caption{Key ideas underpinning our semi-classical model. (a) Numerical simulations of the (disorder-free) 1D Schr{\"o}dinger equation for $U = 1500\,E_0$ and $\omega = 75\,E_0/\hbar$, starting from different initial sine-basis states (red dots).
    The density plots show the 1D momentum distribution $n^{\rm 1D}_k(k_z, t)$.
    The horizontal dashed line indicates the cutoff momentum $\kc$ (see text and Appendix \ref{sec:kc-calculation}).
    (b) The unbounded energy growth process in our model. The dots indicate the sine-basis eigenstates $|\mathbf{k}\rangle$ of the disorder-free box. At $t = 0$, the particles start in the ground state (red dot), and their $k_z$ may be increased by the drive (blue arrows) up to $\kc$. The disorder-induced scattering (orange arrows) moves the particles along equal-energy shells and provides opportunities for the drive to further pump energy into the system.
    }
    \label{fig:2}
\end{figure*}

\section{Semi-classical model}
\label{sec:Semiclassical_model}
The key ideas used to develop our model for the interplay of the drive and disorder are illustrated in Fig.~\ref{fig:2}.
First, we note that in the absence of disorder, strongly driving the gas along a separable axis of the trap leads to 1D chaotic dynamics with bounded energy growth~\cite{Reichl:1986, Lin:1988,Martirosyan:2023}. This is illustrated by the (disorder-free) 1D Schr{\"o}dinger-equation simulations shown in Fig.~\ref{fig:2}(a), where we initialize the system in different sine-basis states $|\psi(t{=}0)\rangle = |k_{z,0}\rangle$ of the box (red dots).
For small $k_{z,0}$, the strong drive mixes the $k_z$ states only up to a cutoff~$\kc$ (horizontal dashed line), while for large $k_{z, 0}$, the drive only weakly perturbs the system.
However, the presence of disorder significantly modifies the picture in 3D. While the drive can only increase $k_z$ up to about $\kc$, the disorder can scatter particles to equal-energy states with lower $k_z$, where the drive can again increase their energy [see Fig.~\ref{fig:2}(b)]. This cooperative process is the key to the unbounded energy growth.

To model this process, we propose the following semi-classical kinetic equation
\begin{equation}
\begin{split}
    \frac{\partial n_k(\mathbf{k}, t)}{\partial t} = s\, |\mathbf{k}|&\left[ - n_k(\mathbf{k}, t) +  \frac{2}{\pi |\mathbf{k}|^2}\int_{|\mathbf{k}| = |\mathbf{k}'|} n_k(\mathbf{k}', t)~{\textrm d}^2 \mathbf{k}' \right] \\
    +\,\Theta(\kc-k_z)\,f &\left[- n_k(\mathbf{k}, t) + \frac{1}{\kc} \int_0^{\kc} n_k(k_x, k_y, k_z', t)~{\rm d}k_z'\right] \, ,
\end{split}
\label{eq:semi-classical-kinetic-eq}
\end{equation}
where $\Theta(k)$ is the Heaviside function, and $s$ and $f$, respectively, characterize 
the rate of the elastic disorder-induced scattering and the rate at which the drive can change the system’s energy.

\begin{figure*}[t]
    \centering
    \includegraphics[width=\textwidth]{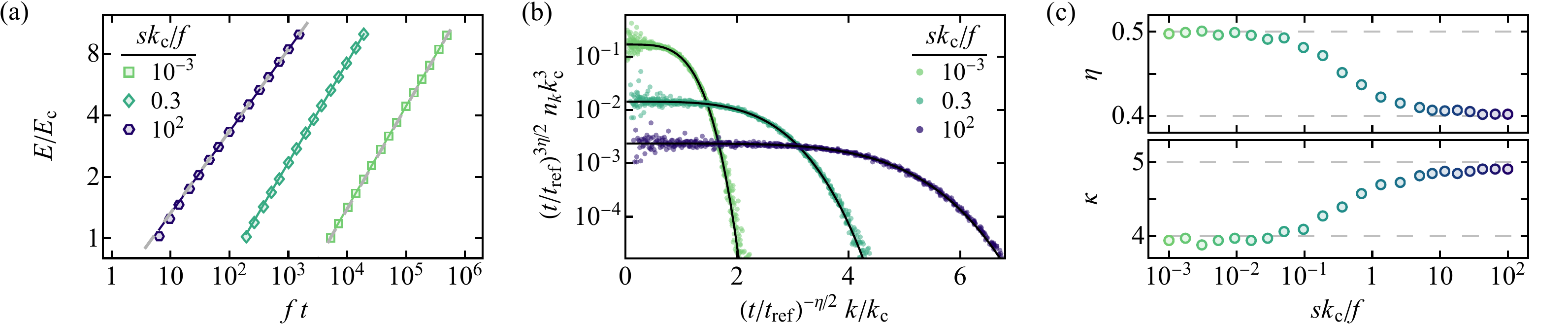}
    \caption{Stochastic-simulation results for our semi-classical model. (a) Energy growth over time for different $s \kc/f$. We normalize the energy $E$ by $E_{\textrm c} = \hbar^2\kc^2/(2m)$, and the time $t$ by $1/f$. The colored lines are power-law fits used to extract $\eta$. The gray dashed lines show
    analytic predictions for $s \kc/f$ = $10^2$ and $10^{-3}$ using Eq.~(\ref{eq:high-scattering-E-t}) and Eq.~(\ref{eq:high-driving-E-t}), respectively. 
    (b) Momentum distributions for the simulations shown in (a), dynamically scaled according to Eq.~(\ref{eq:dynamicscaling}), using the extracted $\eta$ and an arbitrary reference time $t_\text{ref} = 5\times 10^{3}/f$. The solid lines show compressed-exponential fits used to extract $\kappa$.
    (c) Extracted $\eta$ and $\kappa$ as a function $s \kc/f$, showing a similar crossover between the two dynamical regimes as seen in Fig.~\ref{fig:1}(f).
    }
    \label{fig:3}
\end{figure*} 

The first line of Eq.~(\ref{eq:semi-classical-kinetic-eq}) describes the elastic disorder-induced scattering. A perturbative treatment using Fermi's golden rule gives the scattering rate from a state $|\mathbf{k}\rangle$ as
\begin{equation}
    \Gamma_{\rm s}(\mathbf{k})  = \frac{2\pi}{\hbar}\sum_{\mathbf{k}'}  |\langle \mathbf{k} | \VD | \mathbf{k}' \rangle|^2 \, \delta[E(\mathbf{k}) - E(\mathbf{k'})]\,.
    \label{eq:scattering-rate gamma}
\end{equation}
For uncorrelated $\VD$, after ensemble-averaging $|\langle \mathbf{k} | \VD | \mathbf{k}' \rangle|^2$ is $\mathbf{k}$-independent, so
\begin{equation}
    \Gamma_{\rm s}(\mathbf{k}) \simeq \langle \Gamma_{\rm s}(\mathbf{k}) \rangle = s \, |\mathbf{k}| \, ,
    \label{eq:scattering-rate}
\end{equation}
where $s \propto \sigma^2$ (see Appendix~\ref{sec:scattering-rate-calculation}), and the factor of $k$ arises from the 3D density of states. This leads to the $-s\,|\mathbf{k}| n_k(\mathbf{k}, t)$ term in Eq.~(\ref{eq:semi-classical-kinetic-eq}) for the population {\it out-flux} from state $|\mathbf{k}\rangle$. 
The integral term in the first line of Eq.~(\ref{eq:semi-classical-kinetic-eq}) describes the {\it in-flux} to state  $|\mathbf{k}\rangle$; since the out-flux from each state contributes an equal in-flux to every other state on the same $k$-shell, the total in-flux to state $|\mathbf{k}\rangle$ due to scattering is given by the out-flux averaged over the shell.

The second line of Eq.~(\ref{eq:semi-classical-kinetic-eq}) heuristically models the driving process. While the chaotic 1D dynamics is not amenable to an exact treatment, the simulation results in Fig.~\ref{fig:2}(a) inspire a simple model, where the drive randomly mixes $k_z$ states up to $\kc$ at a phenomenological rate $f$, without affecting states with $k_z > \kc$.
While we also treat $\kc$ phenomenologically, its value can be estimated from the time-averaged energy of the driven 1D system (see Appendix \ref{sec:kc-calculation}). 

Before analytically studying this model, we validate it through stochastic numerical simulations of Eq.~(\ref{eq:semi-classical-kinetic-eq}) for different values of the dimensionless parameter $s k_{\rm c} /f$, which sets the ratio of the elastic scattering rate to the rate at which the drive can change the system's energy. As shown in Fig.~\ref{fig:3}, our model, despite its simplicity, captures all the key features seen in the Schrödinger-simulation results in Fig.~\ref{fig:1}.

\section{Analytic analysis}
\label{sec:Analytic_analysis}

\begin{figure*}[t]
    \centering
    \includegraphics[width=1\textwidth]{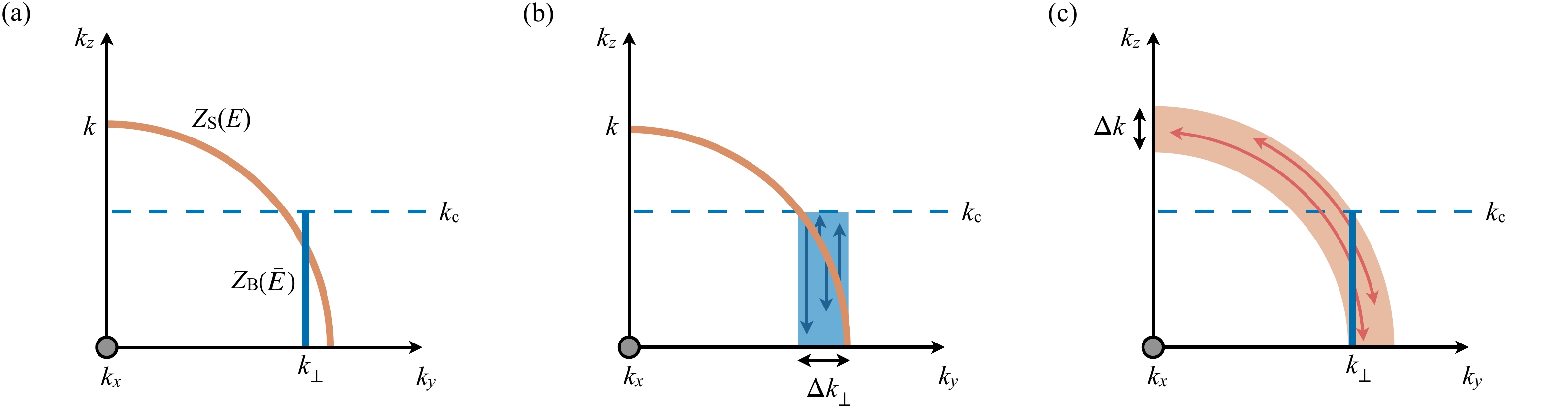}
      \caption{
      Qualitative ideas underpinning the energy-space random-walk picture. (a) Statistical states of the particle in energy space: $Z_\text{S}(E)$ (orange) labels the state of a particle randomly distributed on a spherical shell with definite $k$ and $E$, while $Z_\text{B}(\bar{E})$ (blue) labels the state of a particle randomly distributed in a cylindrical band with definite $k_\perp$, random $k_z\in[0, \kc]$, and  mean energy $\bar{E}$.
      (b) Effects of a driving event after scattering events. When the driving event happens, the particle originally in $Z_\text{S}(E)$ (orange line) may be driven into a range of $Z_\text{B}(\bar{E})$ states (blue shaded area) with a range ($\Delta k_\perp$) of $k_\perp$, and hence a range of $\bar{E}$.
      (c) Effects of a scattering event after driving events. When the scattering event happens, the particle originally in state $Z_\text{B}(\bar{E})$ (vertical blue line) may be scattered into $Z_\text{S}(E)$ states (orange shaded area) with a range ($\Delta k$) of $k$, and hence a range of $E$ .
      }
    \label{fig:4}
\end{figure*}

\subsection{Qualitative ideas}
\label{sec:Qualitative-ideas-and-limiting-regimes}
To solve our model analytically, we switch from momentum space, where Eq.~(\ref{eq:semi-classical-kinetic-eq}) describes a highly non-local jump process, to energy space, where the process is quasi-local.
In energy-space, the trajectory of a particle can be described by a sequence of events of the form
\begin{eqnarray*}
& \text{...}\xrightarrow{S} Z_\text{S}(E_1) \xrightarrow{S} Z_\text{S}(E_1) \xrightarrow{S} Z_\text{S}(E_1) \xrightarrow{D} Z_\text{B}(\bar{E}_2) \\ 
& \xrightarrow{D} Z_\text{B}(\bar{E}_2) \xrightarrow{S} Z_\text{S}(E_3) \xrightarrow{S} Z_\text{S}(E_3) \xrightarrow{D} Z_\text{B}(\bar{E}_4) \xrightarrow{}...\,,
\end{eqnarray*}
where $S$ and $D$ refer to individual scattering and driving events, and $Z_\text{S}(E)$ and $Z_\text{B}(\bar{E})$ label the state of the particle, where the subscripts stand for \emph{shell} and \emph{band}, respectively [see Fig.~\ref{fig:4} (a)].
Immediately after a scattering event $S$, the particle is randomly distributed on a $k$-\emph{shell} of energy $E$, so its state can be labeled $Z_\text{S}(E)$.
Similarly, immediately after a driving event $D$, the particle is randomly distributed on a cylindrical \emph{band} with radius $k_\perp = (k_x^2 + k_y^2)^{1/2}$ and unknown $k_z\in[0, \kc]$, so only its mean energy $\bar{E} = {\hbar^2} \left(k_\perp^2 + \kc^2/3\right)/({2m})$ is known;
we label such a state $Z_\text{B}(\bar{E})$.

Note that successive $S$ or $D$ events do not change the state $Z_\text{S}$ or $Z_\text{B}$, but the state changes when $S$ and $D$ alternate, as illustrated in Figs.~\ref{fig:4}(b) and (c).
When the particle is in $Z_\text{S}(E)$, a $D$ event can drive it into $Z_\text{B}$ states with a range of possible $k_\perp$ [see $\Delta k_\perp$ in Fig.~\ref{fig:4}(b)], and hence a range of $\bar{E}$ distributed around the original energy $E$.
Similarly, when the particle is in $Z_\text{B}(\bar{E})$, an $S$ event can scatter it into $Z_\text{S}$ states  with a range of possible $k$ [see $\Delta k$ in Fig.~\ref{fig:4}(c)], and hence a range of $E$ distributed around the original energy $\bar{E}$.
In both cases, the energy change can be of either sign and has an absolute value on the order of $\Ec = \hbar^2 \kc^2/(2m)$.
This suggests an energy-space random walk with
\begin{equation}
    \frac{\text{d}}{\text{d}t} \left\langle E^2 \right\rangle \propto r(E) \Ec^2 \, ,
    \label{eq:qualitative-random-walk}
\end{equation}
where $r(E)$ is the (generally energy-dependent) rate at which $S$ and $D$ alternate.

In the strong-scattering regime, $sk \gg f$, the rate $r(E)$ is limited by the occurrence of $D$ events, so $r(E) \propto ({\kc}/{k}) \, f$, where
the ${\kc}/{k}$ factor arises because the particle takes part in the random walk only when $k_z < \kc$. Since $k\propto\langle E^2\rangle^{1/4}$, Eq.~(\ref{eq:qualitative-random-walk}) implies
\begin{equation}
    E \propto \Ec \,\left(f t\right)^{{2}/{5}},
    \label{eq:qualitative-E-t-strong-scattering}
\end{equation}
in agreement with both the Schr{\"o}dinger-equation simulations and the stochastic simulations of our model.

On the other hand, in the strong-driving regime, $f \gg sk$, the rate $r(E)$ is limited by the occurrence of $S$ events and given by $r(E) \propto ({\kc}/{k}) \, sk = s \kc$.
As the suppression from  $\kc/k$ is cancelled by the density of states factor $k$ in the scattering rate, Eq.~(\ref{eq:qualitative-random-walk}) implies 
\begin{equation}
    E \propto \Ec \,({s\kc t})^{1/2},
    \label{eq:qualitative-E-t-strong-drive}
\end{equation}
which is also consistent with both the Schr{\"o}dinger-equation simulations and the stochastic simulations of our model.

\begin{figure*}
    \centering
    \includegraphics[width=\textwidth]{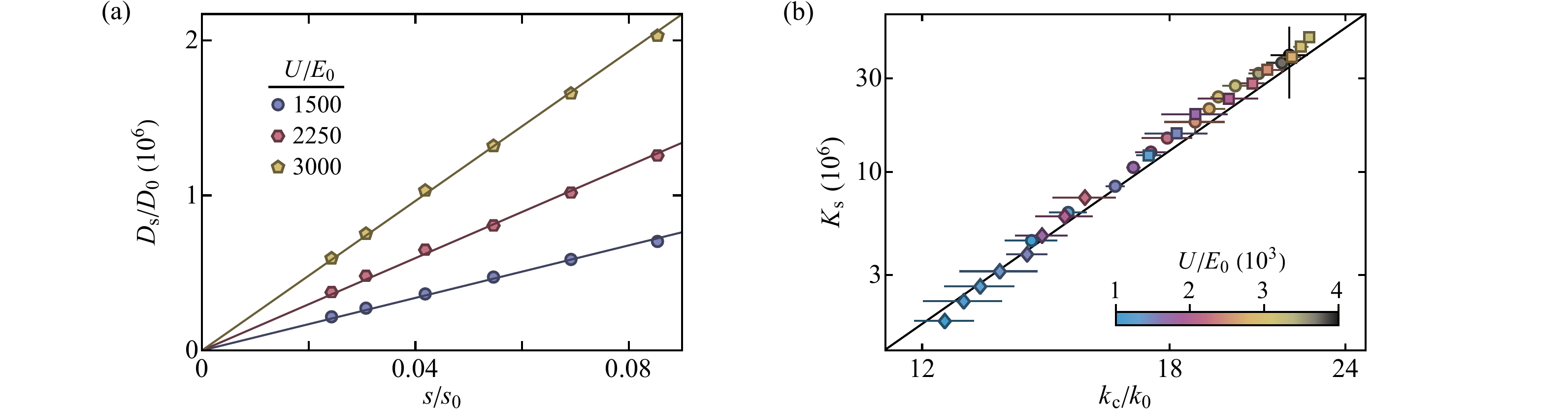}
    \caption{
    Comparison between analytic predictions and Schr{\"o}dinger-equation simulations in the low-disorder regime ($sk\ll f$). (a)
    Energy-space diffusion coefficient $\Ds$ (in units of $D_0 = E_0^3/\hbar$) extracted from simulations with $\omega = 75\,E_0/\hbar$, plotted versus $s$ (in units of $s_0 =E_0 L /\hbar$) analytically calculated from $\sigma$. The solid lines show the linear fits to $\Ds/D_0 =\Ks s/s_0$, used to extract the proportionality constant $\Ks$. (b)~$\Ks$~versus calculated $\kc$ (see Appendix~\ref{sec:kc-calculation}) for different $U$ (color bar) and $\omega$ (diamonds: $50\,E_0/\hbar$, circles: $75\,E_0/\hbar$, and squares: $100\,E_0/\hbar$). The solid line shows the analytic prediction $\Ks=(\pi \kc/k_0)^5/45$ [calculated from Eq.~(\ref{eq:Ds})] with no free parameters.}
    \label{fig:5}
\end{figure*}

\subsection{Energy-space drift-diffusion equation}
\label{sec:Derivation-of-the-energy-diffusion-equation}

We now formalize the ideas from Eq.~(\ref{eq:qualitative-random-walk}) and derive an energy-space drift-diffusion equation valid for all values of $s\kc/f$. Since the particle changes state only when $S$ and $D$ alternate, we can more succinctly describe its trajectory as 
\[...\xrightarrow{\mathbf{S}} Z_\text{S}(E_1) \xrightarrow{\mathbf{D}} Z_\text{B}(\bar{E}_2) \xrightarrow{\mathbf{S}} Z_\text{S}(E_3) \xrightarrow{\mathbf{D}} ... \, , \]
where $\mathbf{S}$ and $\mathbf{D}$, respectively, stand for $D...DS$ and $S...SD$. Heuristically, the energy change may be described by
\begin{equation}
    \frac{\text{d} E}{ \text{d} t} = v_E + \zeta(t) \, ,
    \label{eq:SDE}
\end{equation}
where the `velocity' $v_E$ and `random force' $\zeta(t)$, respectively, lead to energy drift and diffusion.
Denoting $\langle T_{S,D} \rangle$ as the mean waiting time for $\mathbf{S}$ and $\mathbf{D}$ to happen, and $\mu_{S,D}$ and $\sigma^2_{S,D}$ the energy drift and the energy-variance production in each step, we have
\begin{equation}
\setlength{\jot}{5pt}
\begin{split}
    v_E& =  \frac{\mu_S + \mu_D}{\langle T_S + T_D\rangle} \, , \\
    \langle \zeta(t) \zeta(t') \rangle & = \frac{\sigma_S^2 + \sigma_D^2}{\langle T_S + T_D\rangle} \delta(t - t') \, .
\end{split}
\end{equation}
As derived in Appendix \ref{sec:Detailed-derivations-for-the-energy-diffusion-equation}, we have
\begin{equation}
\setlength{\jot}{5pt}
\begin{split}
    \langle T_S + T_D \rangle & = \frac{k}{\kc} \left(\frac{1}{s k} + \frac{1}{f}\right) \, , \\
    \mu_S & = \frac{2 f \Ec^2}{45\, (sk + f)E} \, ,\\
    \mu_D & = 0 \, , \\
    \sigma_{S}^2 & = \sigma_{D}^2 = \frac{4 \Ec^2}{45} \, .
\end{split}
\label{eq:SDE-params}
\end{equation}
The expressions for $\langle T_S + T_D \rangle$, $\mu_D$, and $\sigma_{S,D}$ agree with the qualitative discussion in Section~\ref{sec:Qualitative-ideas-and-limiting-regimes}. The non-zero $\mu_S$ arises because a particle in $Z_\text{B}(\bar{E})$ is more likely to be scattered out of the band at higher $k_z$, due to the $k$ dependence of the scattering rate. 

Combining Eqs.~(\ref{eq:SDE})\,-\,(\ref{eq:SDE-params}), we can write a drift-diffusion equation~\cite{Gardiner:1985} for the energy distribution $P(E, t)$:
\begin{equation}
    \frac{\partial P}{\partial t} = \frac{4 s f \kc \Ec^2}{45}   \frac{\partial}{\partial E}\left[ \frac{1}{s k + f} \left(\frac{\partial P}{\partial E} - \frac{P}{2E} \right)\right]\,.
    \label{eq:generic-diffusion}
\end{equation}
The formal derivation of Eq.~(\ref{eq:generic-diffusion}), using the theory of continuous-time random walks~\cite{Klafter:2011}, is given in Appendix \ref{sec:Detailed-derivations-for-the-energy-diffusion-equation}. This equation satisfies the non-equilibrium fluctuation-dissipation relation proposed in Ref.~\cite{Bunin:2011}, as discussed in Appendix~\ref{sec:neq-fdt}.

\subsection{Limiting regimes}
For $sk \gg f$, Eq.~(\ref{eq:generic-diffusion}) reduces to
\begin{equation}
    \frac{\partial P}{\partial t} = \Dd \frac{\partial}{\partial E}\left[ \frac{1}{\sqrt{E}} \left(\frac{\partial P}{\partial E} - \frac{P}{2E} \right)\right],
    \label{eq:high-scattering-diffusion-equation}
\end{equation}
with diffusion constant 
\begin{equation}
    \Dd = \frac{4}{45}{f\Ec^{5/2}}\,.
\end{equation}
Following Ref.~\cite{Fa:2003}, Eq.~(\ref{eq:high-scattering-diffusion-equation}) can be shown to support self-similar solutions of the form
\begin{equation}
    P(E, t) \propto \frac{E^{1/2}}{(\Dd t) ^{{3}/{5}}} \exp\left[ -\frac{4 E^{{5}/{2}}}{25 \Dd t}\right].
\end{equation}
The corresponding momentum distribution,
\begin{equation}
    n_k(k, t) \propto \frac{1}{(\Dd t) ^{{3}/{5}}} \exp\left[ -\frac{4 \hbar^5 k^5}{25 (2m)^{{5}/{2}} \Dd t}\right] \, ,
    \label{eq:high-scattering-momentum}
\end{equation}
is a compressed exponential [see Eq.~(\ref{eq:compressed-exponential-def})] with $\kappa = 5$, and the energy growth is a power law
\begin{equation}
    \langle E (t) \rangle = \left(\frac{5}{2}\right)^{4/5}\frac{1}{\Gamma(3/5)}\, (\Dd t)^{{2}/{5}} = 1.398\, (\Dd t)^{{2}/{5}}
    \label{eq:high-scattering-E-t} \, ,
\end{equation}
with $\eta = 2/5$ in agreement with Eq.~(\ref{eq:qualitative-E-t-strong-scattering}).

For $sk \ll f$, Eq.~(\ref{eq:generic-diffusion}) reduces to
\begin{equation}
    \frac{\partial P}{\partial t} = \Ds \frac{\partial}{\partial E}\left[ \frac{\partial P}{\partial E} - \frac{P}{2E} \right]\, ,
    \label{eq:high-driving-diffusion-equation}
\end{equation}
with diffusion constant 
\begin{equation}
    D_\text{s} = \frac{4}{45}{s \kc \Ec^2}\,.
    \label{eq:Ds}
\end{equation}
The self-similar solution supported by Eq.~(\ref{eq:high-driving-diffusion-equation}) is 
\begin{equation}
    P(E, t) \propto \frac{E^{1/2}}{(\Ds t)^{3/4}} \exp\left[{-\frac{E^2}{4 \Ds t}}\right].
\end{equation}
The corresponding momentum distribution,
\begin{equation}
    n_k(k, t) \propto \frac{1}{(\Ds t)^{3/4}} \exp\left[{-\frac{\hbar^4 k^4}{16 m^2\Ds t}}\right] \, ,
    \label{eq:high-driving-momentum}
\end{equation}
is a compressed exponential with $\kappa = 4$, and the energy growth is a power law
\begin{equation}
    \langle E (t) \rangle = \frac{\Gamma\left(1/4\right)}{2 \Gamma\left(3/4\right)}\, (D_\text{s}t)^{{1}/{2}} = 1.479 \, (D_\text{s}t)^{{1}/{2}}
    \label{eq:high-driving-E-t}\, ,
\end{equation}
with  $\eta = 1/2$ in agreement with Eq.~(\ref{eq:qualitative-E-t-strong-drive}).

To quantitatively verify Eqs.~(\ref{eq:high-scattering-E-t}) and (\ref{eq:high-driving-E-t}), in Fig.~\ref{fig:3}(a), we compare them
to our stochastic-simulation results for $s\kc/f$ = $10^2$ and $10^{-3}$ and observe good agreement.

In the low-disorder limit ($sk \ll f$), we can also directly compare our analytically predicted energy-diffusion coefficient $\Ds$ [Eq.~(\ref{eq:Ds})] with the Schrödinger-equation simulations, because $\Ds$ depends only on $s$ and $\kc$, both of which can be obtained from the input parameters $\{U$, $\omega$, $\sigma\}$ (such a comparison is not possible for $\Dd$ because we cannot calculate $f$). For each simulation in the low-disorder limit\footnote{We define the low-disorder limit here to correspond to $\eta>0.48$ (as obtained from an unconstrained fit).}, we fit the $E(t)$ curve [such as shown in Fig.~\ref{fig:1}(d)]  to Eq.~(\ref{eq:high-driving-E-t}) and extract $\Ds$. In Fig.~\ref{fig:5}(a), for fixed $\omega$ and various $U$, we plot the extracted $\Ds$ versus $s$ calculated from $\sigma$ and observe the linear behavior predicted in Eq.~(\ref{eq:Ds}). Then, in Fig.~\ref{fig:5}(b), for several $\omega$ and a range of $U$, we show that the fitted constants of proportionality between $\Ds$ and $s$ [the slopes of lines in Fig.~\ref{fig:5}(a)] agree with Eq.~(\ref{eq:Ds}).

\begin{figure*}[t!]
    \centering
        \includegraphics[width=\textwidth, clip]{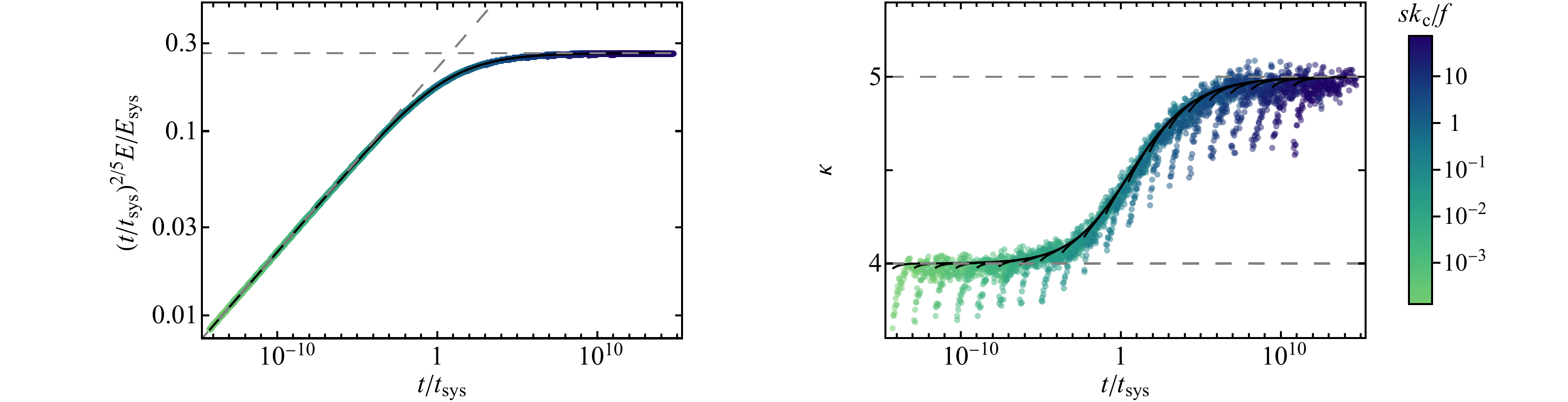}
    \caption{Results of extended stochastic simulations of the semi-classical model with different $s\kc/f$. We show the evolution of the energy $E$~(left) and of the compressed-exponential exponent $\kappa$ for the instantaneous shape of $n_k$~(right).
    We perform simulations with different $s\kc/f$, indicated by the color, which merge into common curves when $t$ and $E$ are scaled using units given in Eq.~(\ref{eq:nd-k-E-t-scale}). 
    In both plots, the dashed lines show the analytic predictions in the two limiting regimes, and the solid black lines show numerical predictions using Eq.~(\ref{eq:generic-diffusion}). 
    The small deviations of stochastic simulations from the solid lines (not visible in the $E$ plot) arise due to initial transients in each simulation.
 }
    \label{fig:6}
\end{figure*}

\subsection{General solution}

After analysing the two limits,  we now examine the general solution to Eq.~(\ref{eq:generic-diffusion}). We can remove the parameters $\{s, \kc, f\}$ from the equation by introducing the dimensionless quantities
\begin{equation}
    k' = \frac{k}{k_\text{sys}}, \quad E' = \frac{E}{E_\text{sys}}, \quad t' = \frac{t}{t_\text{sys}} \, ,
    \label{eq:nd-k-E-t}
\end{equation}
with 
\begin{equation}
    k_\text{sys} = \frac{f}{s}, \quad E_\text{sys} = \frac{\hbar^2 k_\text{sys}^2}{2m }, \quad t_\text{sys} = \frac{f^4}{s^5 \kc^5} \, .
    \label{eq:nd-k-E-t-scale}
\end{equation}
This transforms Eq.~(\ref{eq:generic-diffusion}) to
\begin{equation}
    \frac{\partial P'}{\partial t'} =  \frac{4}{45}  \frac{\partial}{\partial E'}\left[ \frac{1}{k' + 1} \left(\frac{\partial P'}{\partial E'} - \frac{P'}{2E'} \right)\right]\,,
\end{equation}
with $P'(E', t') = E_\text{sys} P(E, t)$. 

This shows that, under appropriate scaling, solutions to Eq.~(\ref{eq:generic-diffusion}) follow a universal $E-t$ trajectory. We illustrate this in Fig.~\ref{fig:6}. In principle, at very long times, one should always observe $E \propto t^{2/5}$ and $\kappa = 5$. 
In terms of system parameters, we classify the system as low-disorder if $s\kc \ll f$ and high-disorder if $s\kc \gg f$, but the dynamics is actually controlled by the ratio $s k/f$, which increases as the energy grows.
Thus, a low-disorder system at long times is mathematically identical to a high-disorder one at short times. However, note that the crossover between the two regimes occurs over an enormous timescale, so any realistic experiment will sample a small region of the universal trajectory, with the energy growth well fitted by a power law and an essentially constant $\kappa$. Also note that for a strongly disordered system, $t_\text{sys}$ is very short ($s\kc/f \gg 1$, so $t_\text{sys} \ll 1/f$), so by the time any significant energy is pumped into the system, $t/t_\text{sys}$ is already large.

\section{Conclusion \& Outlook}

In conclusion, we have developed a semi-classical model for a driven non-interacting box-trapped Bose gas in the presence of uncorrelated disorder. The dynamics at the heart of this model can be understood in terms of an energy-space random walk, and the resulting analytic predictions reproduce the key features seen both in the experiment of Ref.~\cite{Martirosyan:2023} and in Schr{\"o}dinger-equation simulations.

Our work points to several future directions.
First, it would be interesting to experimentally explore the dynamics beyond the weak-disorder regime. This could also lead to further theoretical questions, as the scattering in our model is treated within first-order perturbation theory, which does not hold for arbitrarily strong disorder; for example, the onset of Anderson localization~\cite{Abrahams:2010} may lead to additional dynamical regimes. 
Second, an analogous study in 2D may reveal even richer physics. Taking our model at face value, we would expect $\eta=2/5$ across all parameter regimes in 2D, because there is no density-of-states enhancement factor $k$ in the scattering rate, so we always have $r \propto 1/k$ and $E\propto t^{2/5}$. However, this may be inaccurate due to the more prominent role of fluctuations in 2D. For example, our treatment of the scattering rate relies on ensemble-averaging. While this is a good approximation in 3D, its validity in 2D is not obvious, as far fewer states are involved in the scattering process. This poses interesting questions both experimentally and theoretically.

\section*{Acknowledgements}

The work was supported by EPSRC [Grant No.~EP/P009565/1], ERC (UniFlat), and STFC [Grant No.~ST/T006056/1].
Z.~H. acknowledges support from the Royal Society Wolfson Fellowship.

We dedicate this work to Jean Dalibard, on the occasion of his CNRS Gold Medal.

\clearpage
\onecolumngrid

\renewcommand\thefigure{\thesection\arabic{figure}}    

\appendix
\begin{appendices}
\setcounter{figure}{0}    

\onecolumngrid

\section{Calculation of the semi-classical model parameters}
\subsection{The scattering rate}
\label{sec:scattering-rate-calculation}
In this section, we present the calculation for $\Gamma_{\rm s}(\mathbf{k})$ in Eq.~(\ref{eq:scattering-rate gamma}) and derive an explicit expression for $s$ in Eq.~(\ref{eq:scattering-rate}) for a cubic box of volume $L^3$ in the presence of a disorder potential~$\VD(\mathbf{r})$. The unperturbed basis states of the box are $|\mathbf{k}\rangle$ states of the form
\begin{equation}
    |\mathbf{k}\rangle = \sqrt{\frac{8}{L^3}}\,{\Sigma}(\mathbf{k}, \mathbf{r})= \sqrt{\frac{8}{L^3}} \sin(k_x x)\sin(k_y y)\sin(k_z z) \, .
\end{equation}
The ensemble-averaged matrix element $\llangle[\big] |\langle\mathbf{k}| \VD(\mathbf{r})|\mathbf{k}'\rangle|^2 \rrangle[\big]$ is explicitly
\begin{equation}
\begin{split}
    \llangle[\big] |\langle\mathbf{k}| \VD(\mathbf{r})|\mathbf{k}'\rangle|^2 \rrangle[\big] & = \llangle[\Bigg] \left|\frac{8}{L^3}\int \Sigma(\mathbf{k}, \mathbf{r}) \VD(\mathbf{r}) \Sigma(\mathbf{k}', \mathbf{r}) \, {\text{d}}^3\mathbf{r}\right|^2 \rrangle[\Bigg]\\
    & = \frac{64}{L^6} \int \Sigma(\mathbf{k}, \mathbf{r}_1) \Sigma(\mathbf{k}', \mathbf{r}_1) \Sigma(\mathbf{k}, \mathbf{r}_2) \Sigma(\mathbf{k}', \mathbf{r}_2) \llangle \VD(\mathbf{r}_1) \VD(\mathbf{r}_2) \rrangle \, \rm{d}^3\mathbf{r}_1 \rm{d}^3\mathbf{r}_2\\
    & = \frac{64}{L^6} \int \Sigma(\mathbf{k}, \mathbf{r}_1) \Sigma(\mathbf{k}', \mathbf{r}_1) \Sigma(\mathbf{k}, \mathbf{r}_2) \Sigma(\mathbf{k}', \mathbf{r}_2) C(\mathbf{r}_1 - \mathbf{r}_2) \, \rm{d}^3\mathbf{r}_1 \rm{d}^3\mathbf{r}_2 \,,
\end{split}
\label{eq:matrix element raw}
\end{equation}
where
\begin{equation}
    C(\mathbf{r}_1 - \mathbf{r}_2) = \llangle \VD(\mathbf{r}_1) \VD(\mathbf{r}_2) \rrangle.
\end{equation}
 By substituting the Fourier representation
\begin{equation}
    C(\mathbf{r}_1 - \mathbf{r}_2) = \frac{1}{(2\pi)^3} \int \Tilde{C} (\mathbf{q)} e^{i \mathbf{q} \cdot (\mathbf{r}_1 - \mathbf{r}_2)} \, \rm{d}^3\mathbf{q}
\end{equation}
into Eq.~(\ref{eq:matrix element raw}), we get
\begin{equation}
\begin{split}
    \llangle[\big] |\langle\mathbf{k}| \VD(\mathbf{r})|\mathbf{k}'\rangle|^2 \rrangle[\big] & =\frac{64}{(2\pi)^3 L^6} \int \Sigma(\mathbf{k}, \mathbf{r}_1) \Sigma(\mathbf{k}', \mathbf{r}_1) \Sigma(\mathbf{k}, \mathbf{r}_2) \Sigma(\mathbf{k}', \mathbf{r}_2)\Tilde{C} (\mathbf{q)} e^{i \mathbf{q} \cdot (\mathbf{r}_1 - \mathbf{r}_2)} \, \rm{d}^3\mathbf{r}_1 \rm{d}^3\mathbf{r}_2 \rm{d}^3\mathbf{q} \\
    & = \frac{64}{(2\pi)^3 L^6} \int \left|\int \Sigma(\mathbf{k}, \mathbf{r}) \Sigma(\mathbf{k}', \mathbf{r}) e^{i\mathbf{q}\cdot\mathbf{r}} \, \rm{d}^3\mathbf{r} \right|^2 \Tilde{C} (\mathbf{q)} \, \rm{d}^3\mathbf{q} \, .
\end{split}
\label{eq:matrix element in terms of correlation function}
\end{equation}
The $\{x, y, z\}$ integrals are separable, with 1D integrals of the form
\begin{equation}
\begin{split}
  I_{\rm 1D}&=\left| \int_0 ^L \sin(k x) \sin(k' x) e^{i q x} \, {\rm d}x \right|^2\\
  &= \frac{L^2}{16}\left|\,~\text{sinc}\left(\frac{(k + k' + q)L}{2}\right) e^{i(k+k'+q)L/2} + \text{sinc}\left(\frac{(k + k' - q)L}{2}\right) e^{i(k+k'-q)L/2}\right.  \\
  &\quad\quad - \left.\text{sinc}\left(\frac{(k - k' + q)L}{2}\right) e^{i(k-k'+q)L/2} - \text{sinc}\left(\frac{(k - k' - q)L}{2}\right) e^{i(k-k'-q)L/2}\,\right|^2.
\end{split}
\end{equation}
For large momenta, the sinc functions in the above equation have almost no overlap, so we can drop the interference terms:
\begin{equation}
\begin{split}
  I_{\rm 1D} &\simeq \frac{L^2}{16} \left\{\text{sinc}^2\left(\frac{(k + k' + q)L}{2}\right) + \text{sinc}^2\left(\frac{(k + k' - q)L}{2}\right)\right. + \left.\text{sinc}^2\left(\frac{(k - k' + q)L}{2}\right) +\text{sinc}^2\left(\frac{(k - k' - q)L}{2}\right)\right\}\,,
\end{split}
\end{equation}
and using $\lim\limits_{L\rightarrow \infty}\text{sinc}^2(qL/2) = (2\pi/L)\,\delta(q)$, we get
\begin{equation}
\begin{split}
  I_{\rm 1D} \simeq \frac{L \pi}{8} \left[ \delta (k + k' + q) + \delta (k + k' - q) + \delta (k - k' + q) + \delta (k - k' - q)\right].
\end{split}
\end{equation}
Substituting this expression back into Eq.~(\ref{eq:matrix element in terms of correlation function}), we get
\begin{equation}
    \llangle[\big] |\langle\mathbf{k}| \VD(\mathbf{r})|\mathbf{k}'\rangle|^2 \rrangle[\big] = \frac{1}{64 L^3} \sum_{\mathbf{k}''} \Tilde{C}(\mathbf{k}''),
\label{eq:Vkk}
\end{equation}
with $\mathbf{k}'' = (\lambda_{x} k_x + \lambda_{x}' k_x', \lambda_{y} k_y + \lambda_{y}' k_y', \lambda_{z} k_z + \lambda_{z}' k_z')$ where $\lambda_i, \lambda_i'$ can take on values $ \pm1$. Intuitively, $\{\mathbf{k}''\}$ is the set of $64$ $\mathbf{k}$-vectors that connect the plane-wave components in $|\mathbf{k}\rangle$ to those in $|\mathbf{k}'\rangle$.

In numerical simulations, we discretize real space into $(N-1)^3$ grid points, which requires some changes to the above equations. First, the decomposition of the $C(\mathbf{r}_1-\mathbf{r}_2)$ is written in terms of a Fourier series rather than a Fourier transform,
\begin{equation}
    C(\mathbf{r}_1 - \mathbf{r}_2) = \frac{1}{L^3} \sum_{\mathbf{q}} \Tilde{C}(\mathbf{q}) e^{i\mathbf{q}\cdot(\mathbf{r}_1 -\mathbf{r}_2)}\, ,
\end{equation}
where the summation is performed over the first Brillouin zone of the grid. Second, the delta functions $\delta(\mathbf{k}'' - \mathbf{q})$ become $\sum_{\mathbf{G}}\delta_{\mathbf{k}'' - \mathbf{q}, \mathbf{G}}$ where $\{\mathbf{G}\}$ is the set of reciprocal lattice vectors. This change introduces unphysical Umklapp scattering processes.
However, for the uncorrelated disorder potential $\VD(\mathbf{r})$ with zero mean and variance $\sigma^2$, the correlation function is $C(\mathbf{r}) = \delta_{\mathbf{r}, \mathbf{0}} \sigma^2$, so $\tilde{C}(\mathbf{k}'')$ is constant and the Umklapp scattering processes do not affect the physics.
Eq.~(\ref{eq:Vkk}) then gives
\begin{equation}
    \llangle[\big] |\langle\mathbf{k}| \VD(\mathbf{r})|\mathbf{k}'\rangle|^2 \rrangle[\big] = \frac{\sigma^2}{N^3}\, .
\label{eq:Vkk2}    
\end{equation} 
Note that the factor of $1/64$ in Eq.~(\ref{eq:Vkk}) disappears because all $64$ $\mathbf{k}''$ contribute to the sum.
Finally, substituting Eq.~(\ref{eq:Vkk2}) into Eq.~(\ref{eq:scattering-rate gamma}), we get
\begin{equation}
    \Gamma_{\rm s}(\mathbf{k}) = \frac{2\pi}{\hbar}\sum_{\mathbf{k}'} |\langle\mathbf{k}| \VD(\mathbf{r})|\mathbf{k}'\rangle|^2 \delta[E(\mathbf{k}) - E(\mathbf{k}')] = \frac{m L^3}{\pi N^3 \hbar^3} \sigma^2 k \, ,
\end{equation}
where the sum has been approximated by an integral and we can read off the scattering parameter $s$ in Eq.~(\ref{eq:scattering-rate}) as
\begin{equation}
    s =  \frac{m L^3}{\pi N^3 \hbar^3} \sigma^2 \, .
    \label{eq:s-expression}
\end{equation}

\subsection{The cutoff momentum $\kc$}
\label{sec:kc-calculation}

\begin{figure}[t]
    \centering\includegraphics[width=\textwidth]{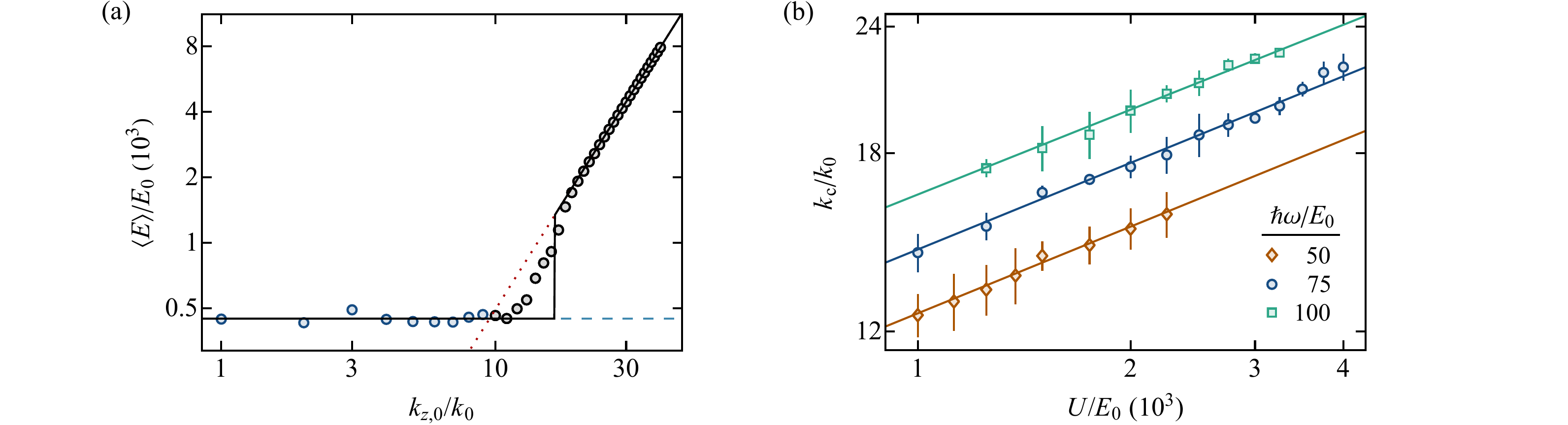}
    \caption{Extraction of $\kc$ from 1D Schrödinger-equation simulations without disorder. 
    (a) Time-averaged energy, $\langle E \rangle$, for the system initialized in different states $k_{z, 0}$; see also Fig.~\ref{fig:2}(a). 
    We normalise $k_{z,0}$ by $k_0 = \pi/L$ and $\langle E \rangle$ by $E_0 = \hbar^2/\left(mL^2\right)$. 
      The solid line shows the prediction of our model: For $k_{z,0} < \kc$, the particles evenly probe momentum states up to $\kc$, so $\langle E \rangle= \Ec/3$ independently of $k_{z, 0}$ (blue dashed line), while for $k_{z,0}>\kc$, we have $\langle E \rangle = \hbar^2 k_{z,0}^2/(2m)$ (red dotted line), so at $k_{z,0} = \kc$ the energy jumps by a factor of $3$.
    (b) Extracted values of $\kc$ for all the simulation parameters used in Fig.~\ref{fig:5}(b). The solid lines are guides to the eye.}
    \label{fig:k_c calculation}
\end{figure}

Here we detail our method to determine $\kc$. 
In our semi-classical model, we have assumed that the drive randomly mixes states with $k_z < \kc$ at a rate $f$. Therefore, in 1D, a state initialized with $k_{z,0} < \kc$ will reach a momentum distribution $n_k^{\rm 1D}$ that is (on average) uniform below $\kc$ and zero above it. For $k_{z,0}<\kc$, the mean energy of the driven system is thus
\begin{equation}
    \langle E \rangle = \frac{1}{\kc} \int_0^{\kc} \frac{\hbar^2}{2m}k_z^2 {\text{d} }k_z = \frac{\hbar^2}{6m} \kc^2 = \frac{1}{3} \Ec \, .
\end{equation}
Therefore, $\kc$ may be estimated by computing $\langle E \rangle$ from 1D Schrödinger-equation simulations for a driven particle in a disorder-free box.
In Fig.~\ref{fig:k_c calculation}(a), we show $\langle E \rangle$ for $U = 1500\,E_0$ and $\omega = 75\,E_0/\hbar$, starting from different $k_{z,0}$. For low $k_{z,0}$, $\langle E \rangle$ is indeed essentially independent of $k_{z,0}$ [see also Fig.~\ref{fig:2}(a)].
To estimate $\kc$ and its error, we use the mean and the standard deviation of $\langle E \rangle$ for $k_{z,0} < 10\,k_0$.
Fig.~\ref{fig:k_c calculation}(b) shows the values of $\kc$ calculated for the values of $U$ and $\omega$ used in Fig.~\ref{fig:5}(b).

\section{Derivation of the energy drift-diffusion equation}
\label{sec:Detailed-derivations-for-the-energy-diffusion-equation}

As described in Section \ref{sec:Derivation-of-the-energy-diffusion-equation}, the trajectory of a particle whose distribution is described by Eq.~(\ref{eq:semi-classical-kinetic-eq}) can be summarized as 
\begin{equation}
    \text{...} \xrightarrow{\textbf{S}} Z_\text{S}(E_1) \xrightarrow{\textbf{D}} Z_\text{B}(\bar{E}_2) \xrightarrow{\textbf{S}} Z_\text{S}(E_3) \xrightarrow{\textbf{D}} Z_\text{B}(\bar{E}_4) \xrightarrow{\textbf{S}} Z_\text{S}(E_5) \xrightarrow{\textbf{D}}... \, ,
\end{equation}
where $\textbf{S}$ and $\textbf{D}$ refer to $D...DS$ and $S...SD$, respectively. In this section, we first calculate the distributions of the waiting times $T_{S,D}$, the energy drifts $\mu_{S,D}(E)$, and the energy variances $\sigma^2_{S,D}(E)$, before assembling the drift-diffusion equation [Eq.~(\ref{eq:generic-diffusion})]. Note that in this section we set $\hbar = m = L = 1$.

\subsection{Calculations for $\textbf{D} = S...SD$}
\subsubsection{Calculation of $T_D$}
The waiting time $T_D$ is a continuous random variable.
We begin by calculating its mean, $\langle T_D \rangle$, before calculating its full distribution.
Suppose that at $t=0$, the particle has just been scattered into $Z_\text{S}(E)$ with unknown position on the $k$-shell. Then, with probability $(1 - \kc/k)$, it sits in the upper part of the shell (`upper shell'), where $k_z > \kc$. In this case, the next event has to be $S$, and the waiting time for it is $\text{Exp}[sk]$ [an exponential distribution with time constant $1/(sk)$]. After the $S$ event, the clock is reset because the particle is still in the same statistical state $Z_{\rm S}$. Therefore, the additional waiting time before $D$ happens is again $T_D$.
Hence,
\begin{equation}
    T_{D|u} = \text{Exp}[sk] + T_D \, ,
    \label{eq:T_D|u}
\end{equation}
where $T_{D|u}$ is the conditional waiting time till $D$ if the particle is initially in the upper shell. On the other hand, if the particle is initially in the lower shell, where $k_z < \kc$, the next event could be either $S$ or $D$, and the waiting time for this event is $\text{Exp}[sk + f]$. With probability $f/(sk + f)$, the event is $D$, and the particle is driven out of the shell. Otherwise, the event is $S$, the clock is reset, and the additional waiting time before $D$ happens is again $T_D$. This can be written as
\begin{equation}
\begin{split}
    T_{D|\ell,D} &= \text{Exp}[sk + f] \, ,\\
    T_{D|\ell,S} &= \text{Exp}[sk + f] + T_D \, , 
\end{split}
\label{eq:T_D|l}
\end{equation}
where $T_{D|\ell,S}$ and $T_{D|\ell,D}$ are the conditional waiting times till $D$ if the particle is initially in the lower shell and the first event after $t = 0$ is $S$ or $D$, respectively. We can thus write an equation for $\langle T_D \rangle$:
\begin{equation}
    \langle T_D \rangle = \frac{k - \kc}{k}\left(\frac{1}{sk} + \langle T_D \rangle \right) + \frac{\kc}{k}\left\{ \frac{f}{sk + f}\frac{1}{sk+f} + \frac{sk}{sk + f} \left( \frac{1}{sk+f} + \langle T_D \rangle\right)\right\}.
\end{equation}
The solution for $\langle T_D \rangle$ is
\begin{equation}
    \langle T_D \rangle = \frac{k}{\kc} \frac{1}{f} + \frac{k - \kc}{\kc} \frac{1}{s k} \, .
\end{equation}
For $k\gg\kc$, to leading order in $\kc$,
\begin{equation}
    \langle T_D \rangle   \simeq \frac{k}{\kc} \frac{1}{f} +  \frac{1}{s \kc} \, .
    \label{eq:TD-leading}
\end{equation}

Generalizing the analysis above,
we can also write an equation for the distribution function $\phi_D(t)$ for $T_D$:
\begin{equation}
    \phi_D(t) = \frac{k - \kc}{k} \int_0^t g(t - t';sk)\, \phi_D(t') \, \text{d} t'+ \frac{\kc}{k} \left(\frac{f}{sk + f} \,g(t;sk + f) + \frac{sk}{sk + f} \int_0^t g(t - t';sk+f)\, \phi_D(t') \, \text{d}t'\right) \, ,    \label{eq:int-eq-phi-d}
\end{equation}
where $g(t; \lambda)$ denotes the exponential distribution function with time constant $1/\lambda$. By taking the Laplace transform
\begin{equation}
    \Tilde{\phi}_D(u) = \mathcal{L}[\phi_D] = \int_0^\infty \phi_D(t) \, e^{-ut}\,  \text{d}t \, ,
\end{equation}
the integral equation Eq.~(\ref{eq:int-eq-phi-d}) can be reduced to an algebraic one, 
\begin{equation}
    \Tilde{\phi}_D(u) =  \frac{k - \kc}{k} \frac{sk}{u + sk}\, \Tilde{\phi}_D(u) + \frac{\kc}{k} \left(\frac{f}{sk + f} \frac{sk + f}{u + sk + f} + \frac{sk}{sk + f} \frac{sk + f}{u + sk + f} \, \Tilde{\phi}_D(u)\right) \, .
\end{equation}
Its solution is
\begin{equation}
    \Tilde{\phi}_D(u) = \frac{\kc}{k} \frac{f u + s f k}{u^2 + (sk +f) u + f s \kc} \, .
\end{equation}
The exact expression for $\phi_D(t)$, obtained from the inverse Laplace transform of $\Tilde{\phi}_D (u)$, is complicated, but it can be well approximated by an exponential distribution. This can be seen from the above equation, where for small $u$ (corresponding to large $t$ and large $k$), we have [using $\langle T_D\rangle$ in Eq.~(\ref{eq:TD-leading})]
\begin{equation}
    \Tilde{\phi}_D(u) \simeq  \frac{\kc}{k} \frac{s f k}{(sk + f) u + f s \kc} \simeq \frac{1/\langle T_D \rangle}{u + 1/\langle T_D \rangle} = \mathcal{L}\left[\frac{1}{\langle T_D \rangle} e^{-t/\langle T_D \rangle}\right]\,.
\end{equation}

\subsubsection{Calculation of $\mu_D(E)$ and $\sigma^2_D(E)$}

When $Z_\text{S}(E) \xrightarrow{\textbf{D}} Z_\text{B}(\bar{E})$ happens,
the new energy $\bar{E}$ is a random variable with distribution $G(\bar{E}|E)$. Irrespective of $T_D$, it is equally likely for the particle to be driven from any point on the lower shell. This means that
\begin{equation}
    G(\bar{E}|E)\, \text{d}\bar{E} \propto 2 \pi k^2 \sin \theta \, \text{d}\theta \propto \frac{k\,k_\perp}{\sqrt{k^2 -k_\perp^2}} \, \text{d} k_\perp \, ,
\end{equation}
with $k_\perp \in\left[\sqrt{k^2 - \kc^2}, k\right]$ because the particle is in the lower shell.
After some manipulation, we get
\begin{equation}
    G(\bar{E}|E) \propto \frac{1}{\sqrt{ E - \bar{E} + \kc^2/6}} \, , 
\end{equation}
with $\bar{E} - E \in \left[-\kc^2/3, \kc^2/6\right]$. From this distribution, we calculate
\begin{equation}
\begin{split}
    \mu_D(E) & = \langle \bar{E} - E \rangle= 0 \, , \\
    \sigma^2_D(E) & = \left\langle (\bar{E} - E)^2 \right\rangle = \frac{1}{45} \kc^4 \, .
\end{split}
\end{equation}

\subsection{Calculations for $\textbf{S} = D...DS$}
\subsubsection{Calculation of $T_S$}
The waiting time $T_S$ is calculated along the same lines as $T_D$ above, but we need to also average over the initial $k_z$ in the band (with fixed $k_\perp$) since the scattering rate $sk$ depends on $k_z$ via  $k = \sqrt{k_\perp^2 + k_z^2}$.
This gives
\begin{equation}
    \langle T_S \rangle = \frac{1}{\kc} \int_0^{\kc} \left[ \frac{1}{s k  + f} + \frac{f}{s k + f} \langle T_S \rangle \right] \, \text{d}k_z \, ,
\end{equation}
where the first term in the integral comes from the mean waiting time till the first event (either $D$ or $S$) after $t = 0$, and the second term comes from the additional time needed if the first event is $D$. Solving the equation, we get
\begin{equation}
    \langle T_S \rangle = \left[{\int_0^{\kc} \frac{s k}{s k + f} \text{d}k_z}\right]^{-1} {\int_0^{\kc} \frac{1}{s k + f}\, \text{d}k_z} \simeq \frac{1}{\sqrt{2\bar{E}} s} + O(\kc^4) \, .
\end{equation}
Note that the integrals in the above equation cannot be evaluated analytically, but by treating $\kc$ and $k_z$ as small parameters, we can perform Taylor expansions and obtain the simple result above.

It is also possible to obtain the distribution of $T_S$ using the Laplace transform,
\begin{equation}
    \Tilde{\phi}_S(u) = \left[\int_0^{\kc} \frac{u + s k}{u + sk + f}\, \text{d}k_z\right]^{-1} {\int_0^{\kc} \frac{s k}{ u + s k + f}\, \text{d}k_z} \simeq \frac{1/\langle T_S \rangle}{u + 1/\langle T_S \rangle} + O(\kc^4) \, .
\end{equation}
Therefore, the distribution for $T_S$ is also exponential to leading order.

\subsubsection{Calculation of $\mu_S(\bar{E})$ and $\sigma^2_S(\bar{E})$}

When $Z_\text{B}(\bar{E}) \xrightarrow{\textbf{S}} Z_\text{S}(E)$ happens, the new energy $E$ is a random variable. The exact distribution for $E$ is tricky to calculate because it is correlated with $T_S$. In particular, if the particle is scattered when it has a higher $k_z$ (and hence a higher $k$), $E$ will be larger, and $T_S$ is likely to have been shorter due to the $k$-dependence of the scattering rate.
However, for an approximate calculation, we ignore this correlation and calculate the distribution of $E$ irrespective of $T_S$. The error of this approximation is a higher-order term. 

First, let us calculate the probability $p(k_z|\bar{E})$ that the particle is scattered out at $k_z$, and hence into the shell with $E = (k_\perp^2 + k_z^2)/2$.
If the particle is at $k_z'$ at $t=0$, then, with probability $sk'/(sk'+f)$, the first event after $t=0$ is $S$, and we get a contribution to $p(k_z|\bar{E})$ only if $k_z'=k_z$; otherwise, with probability $f/(sk'+f)$, the first event after $t=0$ is $D$,
and the probability that the particle leaves at $k_z$ in some future $S$ event is  $p(k_z|\bar{E})$.
Averaging over $k_z'$, we get
\begin{equation}
    p(k_z|\bar{E}) = \frac{1}{\kc} \int_0^{\kc} \left [ \frac{sk'}{sk'+f} \delta(k_z-k_z') + \frac{f}{sk' + f} p(k_z|\bar{E}) \right ] \text{d}k_z' \, ,
\end{equation}
where $k = \sqrt{k_\perp^2 + k_z^2}$ and $k' = \sqrt{k_\perp^2 + k_z'^2}$. Solving the equation, we get
\begin{equation}
    p(k_z|\bar{E}) = \left[\int_0^{\kc} \frac{sk'}{sk' + f} \text{d}k_z'\right]^{-1} \frac{sk}{sk + f} \, .
\end{equation}
Since $E=k^2/2$, we can calculate  
\begin{equation}
\begin{split}
    \mu_S(\bar{E}) & = \left\langle E - 
    \bar{E} \right\rangle = \frac{f \kc^4}{90 \bar{E} \left(f + \sqrt{2 \bar{E}} s \right)} + O(\kc^6) \, , \\
    \sigma^2_S(\bar{E}) & = \left\langle \left(E - \bar{E}\right)^2 \right\rangle = \frac{1}{45}\kc^4 + O(\kc^6) \, .
\end{split}
\end{equation}

\subsection{The drift-diffusion equation}
We can now derive Eq.~(\ref{eq:generic-diffusion}) by generalizing the approach of  Ref.~\cite{Klafter:2011}. First, we denote the probabilities of the particle being in states $Z_\text{S}(E)$ and $Z_\text{B}(E)$ by $P_1(E, t)$ and $P_2(E, t)$, respectively. We express the rate of change of $P_1(E, t)$ and $P_2(E, t)$ as
\begin{equation}
\begin{split}
    \frac{\partial P_1(E, t)}{\partial t} & = J_1^+(E, t) - J_1^-(E, t) \, \\
    \frac{\partial P_2(E, t)}{\partial t} & = J_2^+(E, t) - J_2^-(E, t) \, ,
\end{split}
\label{eq:P-J relation}
\end{equation}
where $J^{+}_{1,2}$ and $J^{-}_{1,2}$ are the in- and out-fluxes, respectively. 
Since the state of the particle has to alternate between $Z_S$ and $Z_B$ between steps of the energy-space random walk, we have
\begin{equation}
\begin{split}
    J_1^+(E, t) & = \int G_2(E|E') J_2^-(E', t) \,\text{d}E' \,,\\
    J_2^+(E, t) & = \int G_1(E|E') J_1^-(E', t) \,\text{d}E',
\end{split}
\label{eq:J+- relation}
\end{equation}
where $G_{1,2}(E|E')$ are the energy transition probabilities for $\mathbf{D}$ and $\mathbf{S}$, respectively.
The out-fluxes $J^-_{1,2}$ can be written as
\begin{equation}
\begin{split}
    J_1^- (E, t) & = \phi_1(E, t) P_1(E, 0) + \int_0^t \phi_1(E, t-t') J_1^+(E, t') \,\text{d}t' \, , \\
    J_2^- (E, t) & = \phi_2(E, t) P_2(E, 0) + \int_0^t \phi_2(E, t-t') J_2^+(E, t') \,\text{d}t' \, ,
\end{split}
\label{eq:out flux expression}
\end{equation}
where $\phi_{1,2}(E, t) = \phi_{D, S}(E, t)$ are the waiting-time distributions. The first terms in Eqs.~(\ref{eq:out flux expression}) represent the fluxes contributed by particles 
originally in the states $1,2$ at $t = 0$, and the second terms represent the fluxes contributed by particles that enter the states at $t'$ and leave at $t$.
Using Eq.~(\ref{eq:P-J relation}), we eliminate $J^+_{1,2}$ from Eq.~(\ref{eq:out flux expression}) and get
\begin{equation}
\begin{split}
    J_1^- (E, t) & = \phi_1(E, t) P_1(E, 0) + \int_0^t \phi_1(E, t-t') \frac{\partial P_1(E, t')}{\partial t'} \, \text{d}t' + \int_0^t \phi_1(E, t-t') J_1^-(E, t') \, \text{d}t' \, , \\
     J_2^- (E, t) & = \phi_2(E, t) P_2(E, 0) + \int_0^t \phi_2(E, t-t') \frac{\partial P_2(E, t')}{\partial t'} \, \text{d}t' + \int_0^t \phi_2(E, t-t') J_2^-(E, t') \, \text{d}t' \, .
\end{split}
\end{equation}
These integral equations can be solved using the Laplace transform, which gives
\begin{equation}
\begin{split}
    \Tilde{J}^-_1(E, u) & = u \Tilde{M}_1(E, u) \Tilde{P}_1(E, u) \, ,  \\
    \Tilde{J}^-_2(E, u) & = u \Tilde{M}_2(E, u) \Tilde{P}_2(E, u) \, ,
\end{split}
\label{eq:laplace out-flux}
\end{equation}
with the memory kernels given by
\begin{equation}
   \Tilde{M}_{1,2}(E, u) =  \frac{\Tilde{\phi}_{1,2}(E, u)}{1 - \Tilde{\phi}_{1,2}(E, u)} \, .
\end{equation}
After Laplace-transforming back, Eq.~(\ref{eq:laplace out-flux}) gives
\begin{equation}
\begin{split}
    J^-_1(E, t) & = \frac{\text{d}}{\text{d}t} \int_0^t M_1(E, t - t') P_1(E, t') \,\text{d}t' \, ,\\
    J^-_2(E, t) & = \frac{\text{d}}{\text{d}t} \int_0^t M_2(E, t - t') P_2(E, t') \,\text{d}t' \, .
\end{split}
\end{equation}
Substituting this into Eq.~(\ref{eq:P-J relation}) and  using Eq.~(\ref{eq:J+- relation}) to eliminate $J^+_{1,2}$, we get
\begin{equation}
\begin{split}
    \frac{\partial P_1(E, t)}{\partial t} & = \int\left[ G_2(E|E') \frac{\text{d}}{\text{d}t} \int_0^t M_2(E', t - t') P_2(E', t') \, \text{d}t' \right]\text{d}E' - \frac{\text{d}}{\text{d}t} \int_0^t M_1(E, t - t') P_1(E, t')\, \text{d}t' \, , \\
    \frac{\partial P_2(E, t)}{\partial t} & = \int\left[ G_1(E|E') \frac{\text{d}}{\text{d}t} \int_0^t M_1(E', t - t') P_1(E', t') \, \text{d}t' \right]\text{d}E' - \frac{\text{d}}{\text{d}t} \int_0^t M_2(E, t - t') P_2(E, t')\, \text{d}t' \, .
\end{split}
\end{equation}
Note that $G_{1,2}(E|E')$ is local, so we can perform a Kramer-Moyal expansion \cite{Gardiner:1985} to convert the above equation to a differential equation in $E$, and the results are
\begin{equation}
\begin{split}
    \frac{\partial P_1(E, t)}{\partial t} &=  \frac{\text{d}}{\text{d}t} \int_0^t \left\{-\frac{\partial}{\partial E}\left[D_2^{(1)}(E) M_2(E, t-t') P_2(E, t')\right] + \frac{1}{2}\frac{\partial^2}{\partial E^2} \left[D_2^{(2)}(E) M_2(E, t-t') P_2(E, t')\right] \right\} \text{d}t'\\
    & ~~~+ \frac{\text{d}}{\text{d}t} \int_0^t\left[M_2(E, t-t') P_2(E, t') - M_1(E, t-t') P_1(E, t')\right] \text{d}t' \, ,\\
    \frac{\partial P_2(E, t)}{\partial t} &=  \frac{\text{d}}{\text{d}t} \int_0^t \left\{-\frac{\partial}{\partial E}\left[D_1^{(1)}(E) M_1(E, t-t') P_1(E, t')\right] + \frac{1}{2}\frac{\partial^2}{\partial E^2} \left[D_1^{(2)}(E) M_1(E, t-t') P_1(E, t')\right] \right\} \text{d}t'\\
    & ~~~+ \frac{\text{d}}{\text{d}t} \int_0^t\left[M_1(E, t-t') P_1(E, t') - M_2(E, t-t') P_2(E, t')\right] \text{d}t' \, , \\
\end{split}
\end{equation}
where $D_{1,2}^{(n)}(E) = \int \Delta E^n \, G_{1,2}(E + \Delta E| E) \, \text{d}\Delta E$ are the Kramer-Moyal coefficients. In the current context, $D_{1,2}^{(1)}(E)$ correspond to $\mu_{S,D}(E)$, and $D_{1,2}^{(2)}(E)$ correspond to $\sigma^2_{S,D}(E)$. Since $\phi_{1,2}(E, t)$ is approximately exponential, we also have
\begin{equation}
    M_{1,2}(E, t) \simeq \frac{1}{\tau_{1,2}(E)} \, ,
\end{equation}
where $\tau_{1,2}(E)$ are the mean waiting times $\langle T_{S,D}\rangle$ at energy $E$. Putting this in, we get the following set of Fokker-Planck equations:
\begin{equation}
\begin{split}
    \frac{\partial P_1(E, t)}{\partial t} & = \frac{P_2(E, t)}{\tau_2(E)} - \frac{P_1(E, t)}{\tau_1(E)} - \frac{\partial}{\partial E}\left[\frac{D_2^{(1)}(E)}{\tau_2(E)} P_2(E, t)\right] + \frac{1}{2}\frac{\partial^2}{\partial E^2}\left[\frac{D_2^{(2)}(E)}{\tau_2(E)} P_2(E, t)\right]  \, , \\  
    \frac{\partial P_2(E, t)}{\partial t} & = \frac{P_1(E, t)}{\tau_1(E)} - \frac{P_2(E, t)}{\tau_2(E)} - \frac{\partial}{\partial E}\left[\frac{D_1^{(1)}(E)}{\tau_1(E)} P_1(E, t)\right] + \frac{1}{2}\frac{\partial^2}{\partial E^2}\left[\frac{D_1^{(2)}(E)}{\tau_1(E)} P_1(E, t)\right] \, . 
\end{split}
\label{eq:P1 p2 Fokker Plancks}
\end{equation}
With rapid relaxation~\cite{Gardiner:1985}, we have ${P_2(E, t)}/{\tau_2(E)} - {P_1(E, t)}/{\tau_1(E)} \simeq 0$ and
\begin{equation}
    P_{1,2}(E, t) = \frac{\tau_{1,2}(E)}{\tau_1(E) + \tau_2(E)} P(E, t) \, ,
\end{equation}
with $P(E,t) = P_1(E, t) + P_2(E, t)$. Substituting this into Eq.~(\ref{eq:P1 p2 Fokker Plancks}), we get
\begin{equation}
    \frac{\partial P(E, t)}{\partial t} = - \frac{\partial}{\partial E}\left[\frac{D_1^{(1)}(E) + D_2^{(1)}(E)}{\tau_1(E) + \tau_2(E)} P(E, t)\right] + \frac{1}{2}\frac{\partial^2}{\partial E^2}\left[\frac{D_1^{(2)}(E) + D_2^{(2)}(E)}{\tau_1(E) + \tau_2(E)} P(E, t)\right] \, .
\end{equation}
We finally get
\begin{equation}
    \frac{\partial P(E, t)}{\partial t} = \frac{s f \kc^5}{45} \frac{\partial}{\partial E} \left[ \frac{\partial}{\partial E}\left(\frac{1}{sk + f} P\right) - \frac{f}{2 (sk + f)^2 E} P \right],
    \label{eq:diffusion-A-B}
\end{equation}
or, after some manipulation,
\begin{equation}
    \frac{\partial P(E, t)}{\partial t} = \frac{s f \kc^5}{45} \frac{\partial}{\partial E}\left[ \frac{1}{s k + f} \left(\frac{\partial P}{\partial E} - \frac{P}{2E} \right)\right],
\end{equation}
which is Eq.~(\ref{eq:generic-diffusion}).

\section{Non-equilibrium fluctuation-dissipation relation}
\label{sec:neq-fdt}

In cases where the dynamics of a periodically driven and thermally isolated system obeys an energy-space drift-diffusion equation of the form
\begin{equation}
    \frac{\partial P}{\partial t} = -\frac{\partial}{\partial E} \left[A(E) P\right] + \frac{1}{2} \frac{\partial^2}{\partial E^2} \left[B(E) P\right],
\end{equation}
Ref.~\cite{Bunin:2011} proposed a general relation between the system-dependent drift and diffusion coefficients [$A(E)$ and $B(E)$, respectively],
\begin{equation}
    2 A(E) = \mathrm{\beta}_T(E) B(E)  + \frac{\partial}{\partial E}B(E),
    \label{eq:neq-fdt}
\end{equation}
where $\beta_T(E) = \partial_E \ln\Omega(E)$ is the micro-canonical inverse temperature defined via the density of states $\Omega(E)$.
This relation is the non-equilibrium version of the equilibrium fluctuation-dissipation theorems, but its range of validity is not well established.

In our case, $\Omega(E)\sim\sqrt{E}$ and $\beta_T = 1/(2E)$. Reading off $A(E)$ and $B(E)$ from Eq.~(\ref{eq:diffusion-A-B}) shows that Eq.~(\ref{eq:neq-fdt}) is satisfied by our drift-diffusion equation in all regimes. This is not surprising because our drift-diffusion equation is derived from a semi-classical kinetic equation [Eq.~(\ref{eq:semi-classical-kinetic-eq})] with reciprocal transition probabilities, $\mathcal{T}(\mathbf{k}\xrightarrow{}\mathbf{k}') = \mathcal{T}(\mathbf{k}'\xrightarrow{}\mathbf{k})$. This reciprocity implies that a system with equal occupation in every state [a uniform $n_k(\mathbf{k})$] must be a stationary state. Correspondingly, when $P(E)\sim\Omega(E)$, the probability current $J(E) = A(E) P(E) - (1/2)~\partial_E[B(E) P(E)]$ should be zero, and this implies Eq.~(\ref{eq:neq-fdt}) as discussed in Ref.~\cite{Bunin:2011}.

\end{appendices}

\end{document}